\documentclass[showpacs,aps,prd,floatfix,amsmath,amssymb,nofootinbib,superscriptaddress]{revtex4-2}
\usepackage{graphicx}
\usepackage{amssymb}
\usepackage{bbm}
\usepackage{hyperref}
\usepackage{color}
\usepackage{slashed}
\usepackage{float}
\usepackage{subfigure}
\usepackage{dcolumn}
\usepackage{hyperref}
\usepackage{physics}

\usepackage{multirow}
\begin{document}

\title{Light quark contributions to Higgs decays}
\author{A. I. Hern\'andez-Ju\'arez}
\email{alan.hernandez@cuautitlan.unam.mx}
\affiliation{Departamento de F\'isica, FES-Cuautitl\'an, Universidad Nacional Aut\'onoma de M\'exico, C.P. 54770, Estado de M\'exico, M\'exico.}
\author{R. Gait\'an}
\affiliation{Departamento de F\'isica, FES-Cuautitl\'an, Universidad Nacional Aut\'onoma de M\'exico, C.P. 54770, Estado de M\'exico, M\'exico.}
\author{ R. Martinez}
\affiliation{Departamento de F\'isica, Universidad Nacional de Colombia, K. 45 No. 26-85, Bogot\'a D.C., Colombia}
\date{\today}

\date{\today}

\begin{abstract}
The literature establishes that the light fermions contributions to the decays $H\to Z\gamma$ and $H\to\gamma\gamma$ are negligible since their coupling with the Higgs is proportional to $m_f$. In the present letter, we show that although such a conclusion is true for leptons, the light quark contributions are zero when we consider their non-perturbative effects.

\end{abstract}


\date{\today}

\maketitle
\section{ Introduction}
\label{intro}

Precision measurements of Higgs couplings at the LHC are essential for assessing whether they align with the predictions of the SM. Recently, the CMS and ATLAS collaborations announced the first measurements of the Higgs boson width, as well as the off-shell contributions of the Higgs to the production of two $Z$ bosons \cite{CMS:2022ley, ATLAS:2023dnm}. Additionally, both collaborations have observed an excess in the production of $Z\gamma$ pairs through a Higgs boson \cite{CMS:2022ahq, ATLAS:2023yqk}.

\begin{figure}[H]
\begin{center}
\subfigure[]{\includegraphics[width=5cm]{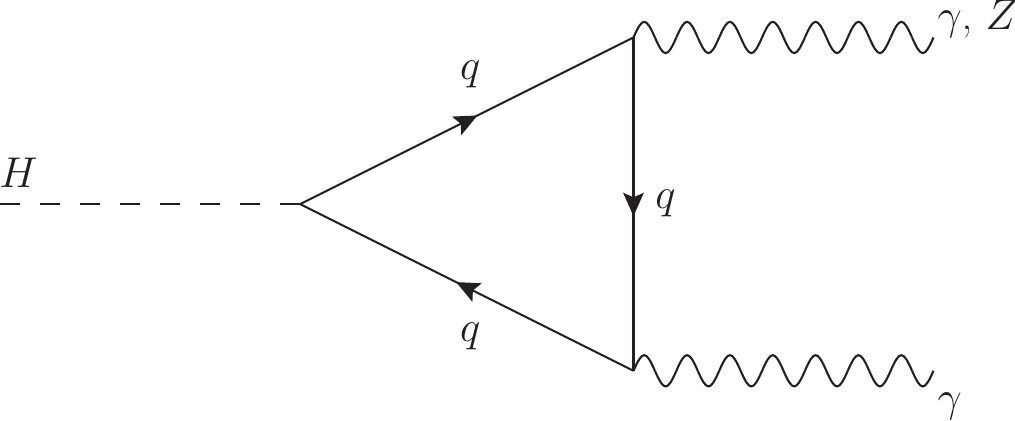}\label{diag3}}
\subfigure[]{\includegraphics[width=5cm]{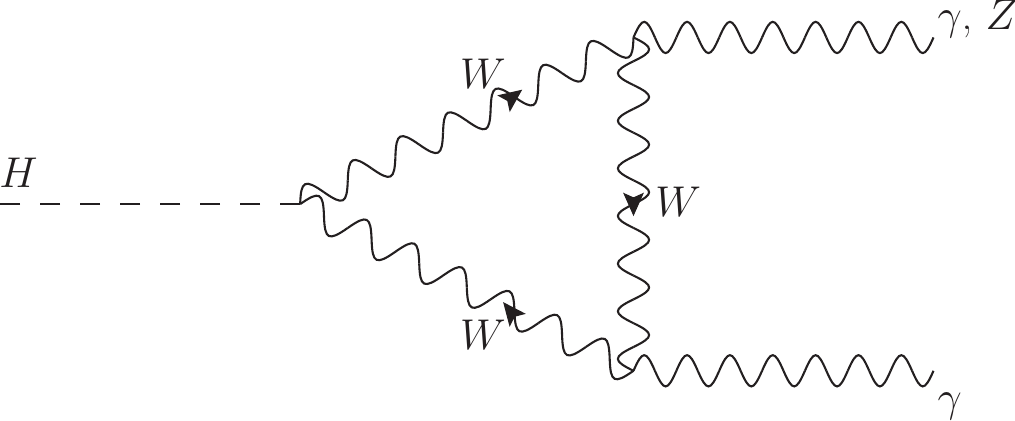}\label{diag1}}
\subfigure[]{\includegraphics[width=5cm]{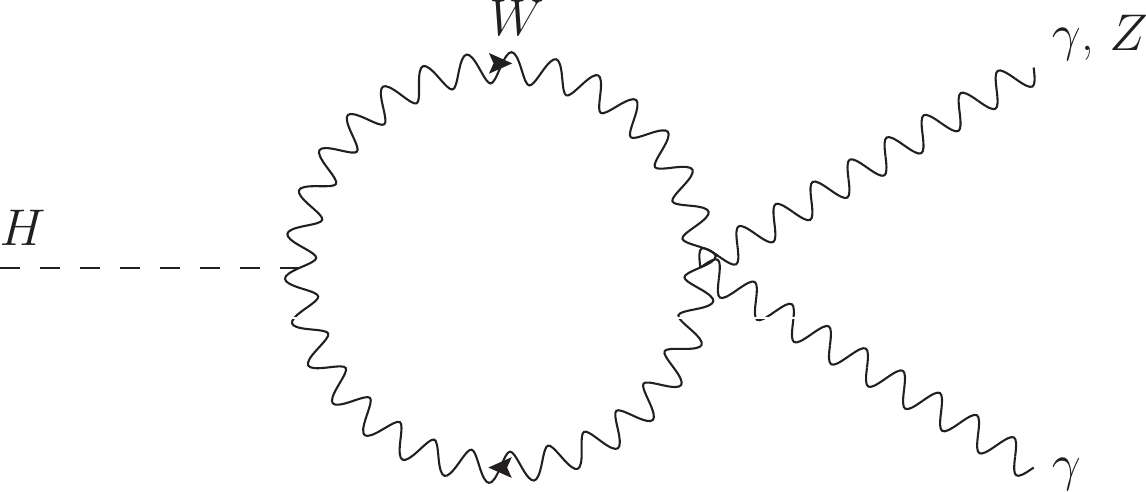}\label{diag2}}
\caption{One-loop SM contributions in the unitary gauge to the $H\rightarrow Z\gamma$ decay. } \label{SMContributions}
\end{center}
\end{figure}

\begin{figure}[H]
\begin{center}
\includegraphics[width=8cm]{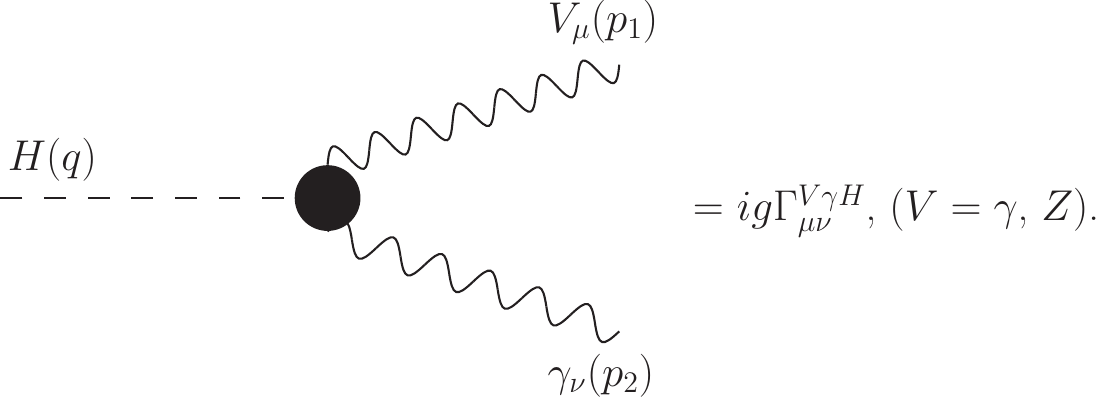}
\caption{$HZ\gamma$ coupling} \label{vertex}
\end{center}
\end{figure}

The $H\to \gamma\gamma$ and $H\to Z\gamma$ decays are loop-induced in the SM through the Feynman diagrams in Fig. \ref{SMContributions}, with the vertex function for both processes given as follows \cite{MARTINEZ1990503, Martinez:1989bg}:  
\begin{equation}
\label{VertexFunction}
\Gamma_{V\gamma H}^{\mu\nu}=h_1^{V\gamma} g^{\mu\nu}+\frac{1}{m_Z^2}\Big\{ h^{V\gamma}_2 p_1^\nu p_2^\mu+h^{V\gamma}_3\epsilon^{\mu\nu\alpha\beta}p_{1\alpha}p_{2\beta}\Big\},\quad (V=\gamma\text{, }Z),
\end{equation}
where we follow the notation in Fig \ref{vertex}. The $h_{1,2}^{V\gamma}$ form factors are $CP$-conserving, whereas $h_3^{V\gamma}$ is associated with $CP$ violation. Due to gauge invariance, the $h^{V\gamma}_1$ and $h^{V\gamma}_2$ form factors are not independent of each other:
\begin{equation}\label{h2toh1}
h^{Z\gamma}_2=\frac{2\  m^2_Z}{m_Z^2-m_H^2}h^{Z\gamma}_1,
\end{equation}
\begin{equation}
h_2^{\gamma\gamma}=-2\ h_1^{\gamma\gamma}.
\end{equation}

The $CP$-conserving form factors are induced within the SM at the one-loop level, and these were calculated long ago for the $H \gamma\gamma$ \cite{ELLIS1976292, Shifman:1979eb, Okun1982LeptonsAQ, GAVELA1981257} and $H Z\gamma$ \cite{Cahn:1978nz, Bergstrom:1985hp, Gunion:1989we, Decker:1991dz} vertices. The most significant contributions come from the $W$-loop diagrams shown in Figs. \ref{diag1}-\ref{diag2}, and are given as \cite{Hernandez-Juarez:2024pty}:
\begin{align}
\label{}
    h_1^{Z\gamma}(W)&=-3.4\times 10^{-1} \text{GeV}   \\
        h_1^{\gamma\gamma}(W)&=4.8\times 10^{-1} \text{GeV}  
\end{align}

In the literature, contributions from heavy fermions are usually emphasized, while those from light quarks and leptons are regarded as negligible. This conclusion arises from the coupling  $H\overline{f}f$, which is proportional to the mass $m_f$. Consequently, the contributions to $h_1^{ZV}$ ($V=Z$, $\gamma$) will be proportional to $m_f^2$ and can be neglected (see the discussion of one-loop contributions of light fermions in Refs \cite{ELLIS1976292, Cahn:1978nz, Bergstrom:1985hp, Decker:1991dz, BhupalDev:2013xol, Spira:1991tj, Bonciani:2015aa, Gehrmann:2015aa, PhysRevD.42.3760, PhysRevD.47.1264, DJOUADI1993255, DJOUADI1991187, Melnikov_1993, FLEISCHER2004294, DJOUADI199817}). Nevertheless,  this argument is based on a perturbative approach, which is inadequate for light quarks due to their non-perturbative effects. 

In this study, we aim to investigate the role of light quarks in the decay processes $H\to Z\gamma$ and $H\to \gamma\gamma$ to determine whether their contributions can be considered insignificant or zero.

 \section{Fermion contributions to $h^{V\gamma}_1$ ($V=Z$, $\gamma$)}
 
 In perturbation theory, the fermion contribution from the diagram in Fig. \ref{diag3} can be expressed in terms of the Passarino-Veltman scalar functions as follows
\begin{align}
\label{h1F}
h_1^{Z\gamma}(F)=&\frac{N_c \mathcal{Q}_f\ g\ g_V\ e\ m_f^2 }{8 \pi ^2  c_W m_W
   \left(m_H^2-m_Z^2\right)} \Bigg\{2
   m_Z^2\Big[
   \text{B}_0\left(m_H^2,m_f^2,m_f^2
   \right)-
   \text{B}_0\left(m_Z^2,m_f^2,m_f^2
   \right)\Big]\nonumber\\
   &+\left(m^2_H-m^2_Z\right)
   \left(\left(-m_H^2+4
   m_f^2+m_Z^2\right)
   \text{C}_0\left(0,m_H^2,m_Z^2,m_f
   ^2,m_f^2,m_f^2\right)+2\right)\Bigg\},
\end{align}
\begin{align}
\label{h1Fgamma}
h_1^{\gamma\gamma}(F)=-\frac{ e^2 \mathcal{Q}_f^2 m_f^2\ N_c }{8 \pi
   ^2 m_W}
   \Big\{2-\left(m_H^2-4
   m_f^2\right)
   \text{C}_0\left(0,0,m_H^2,m_f^2,m
   _f^2,m_f^2\right)\Big\}.
\end{align}
where $m_f$ and $\mathcal{Q}_f$ denote the mass and charge of the fermion $f$ in the loop, respectively. Additionally, $g_V$ corresponds to the vector coupling of the $Z$ boson with the fermions, and $N_c$ stands for the color number. Contributions from heavy quarks and leptons can be calculated using Eqs \eqref{h1F} and \eqref{h1Fgamma}. For this analysis, we adopt the masses $m_t=172.57$ GeV, $m_c=1.51$ GeV, and $m_b=4.18$ GeV, as recommended by the LHCHWG \cite{LHCHiggsCrossSectionWorkingGroup:2016ypw, ParticleDataGroup:2022pth}. Public codes, such as \texttt{HDECAY}, have implemented the masses $\overline{m}_q(\mu)$ in the $\overline{\text{MS}}$ scheme to assess the contributions of the bottom and charm quarks \cite{Djouadi_2019hdec}. The numerical values from the top, bottom, and charm quarks, along with the total contributions from leptons, are summarized in Table. \ref{H1contributions}. 

The imaginary parts of the form factors $h_1^{V\gamma}$ ($V=Z$, $\gamma$) emerge when the fermions in the loop that are coupled to the Higgs boson can be on-shell. Notably, the contributions from the top quark are comparable in magnitude to those arising from the $W$-loops. The remaining contributions, however, are two to three orders of magnitude smaller than the top quark contributions and can be considered negligible. This finding is expected from Eqs \eqref{h1F}-\eqref{h1Fgamma}, where we find that $h_1^{V\gamma}\sim m_f^2$, reinforcing the argument found in the literature to neglect the contributions from light fermions.

\begin{table}[H] \begin{center}\begin{tabular}{ccc} Contribution & $h_1^{\gamma\gamma}$ & $h_1^{Z\gamma}$  \\\hline\hline Top quark & -1.07$\times10^{-1}$ GeV&1.9$\times 10^{-2}$ GeV  \\\hline Bottom quark &$(1.399\times 10^{-3}-i\ 1.84\times10^{-3})$ GeV& $(-3.91\times 10^{-4}+i\ 2.20\times10^{-4})$ GeV   \\\hline Charm quark & $(1.13\times 10^{-3}-i\ 9.26\times10^{-4})$ GeV&$(-8.07\times 10^{-5}+i\ 3.32\times10^{-5})$ GeV    \\\hline Leptons &$(1.395\times 10^{-3}-i\ 1.26\times10^{-3})$ GeV& $(-1.46\times 10^{-5}+i\ 6.3\times10^{-6})$  GeV
 \end{tabular}\caption{Perturbative contributions to the form factors $h_1^{\gamma\gamma}$ and $h_1^{Z\gamma}$ from the top, bottom, charm quarks, and leptons. For the quark masses, we use the values recommended by the LHCHWG \cite{LHCHiggsCrossSectionWorkingGroup:2016ypw, ParticleDataGroup:2022pth}.}\label{H1contributions}\end{center}
 \end{table}

  The contributions from light fermions require a different treatment, as their masses are not well-defined within a perturbative regime. One viable approach is to consider the chiral limit \cite{PhysRevLett.28.1482, LEUTWYLER1974413, Maris_1997}, defined in a non-perturbative framework. In QCD, chiral symmetry is dynamically broken \cite{PhysRevC.68.015203, Bhagwat_2006, Bowman_2005},  which provides a mechanism for generating quark masses. Within context, one can define and compute one-loop corrections of the light quarks, including their chromomagnetic dipole moments \cite{Chang_2011, PhysRevD.95.034041}. In Table \ref{H12contributions}, we present our calculations of the light quark contributions to $h_1^{V\gamma}$. This analysis was conducted through a perturbative approach, assuming that custodial symmetry is preserved. For the light quark masses, we adopt $m_u\approx m_d\approx 8$ MeV and $m_s\approx 150$ MeV for the strange quark. We observe that the contributions from light quarks are negligible compared with those reported in Table \ref{H1contributions}, as pointed out in the literature. 
  
 \begin{table}[H] \begin{center}\begin{tabular}{ccc} Contribution & $h_1^{\gamma\gamma}$ & $h_1^{Z\gamma}$  \\\hline\hline  Up quark &$(2.28\times 10^{-7}-i\ 7.73\times10^{-8})$ GeV& $(-5.38\times 10^{-9}+i\ 9.33\times10^{-10})$  GeV
\\\hline Down quark &$(5.72\times 10^{-8}-i\ 1.93\times10^{-8})$ GeV& $(-4.64\times 10^{-9}+i\ 8.06\times10^{-10})$  GeV
\\\hline Strange quark &$(9.36\times 10^{-6}-i\ 4.73\times10^{-6})$ GeV& $(-1.1\times 10^{-6}+i\ 2.83\times10^{-7})$  GeV
 \end{tabular}\caption{Perturbative contributions to the form factors $h_1^{\gamma\gamma}$ and $h_1^{Z\gamma}$ from the up, down and strange quarks. For the quark masses, we use $m_u=m_d=8$ MeV and $m_s=150$ MeV. }\label{H12contributions}\end{center}
 \end{table}

 \section{Pion contributions}
 
 \begin{figure}[H]
\begin{center}
\includegraphics[width=8cm]{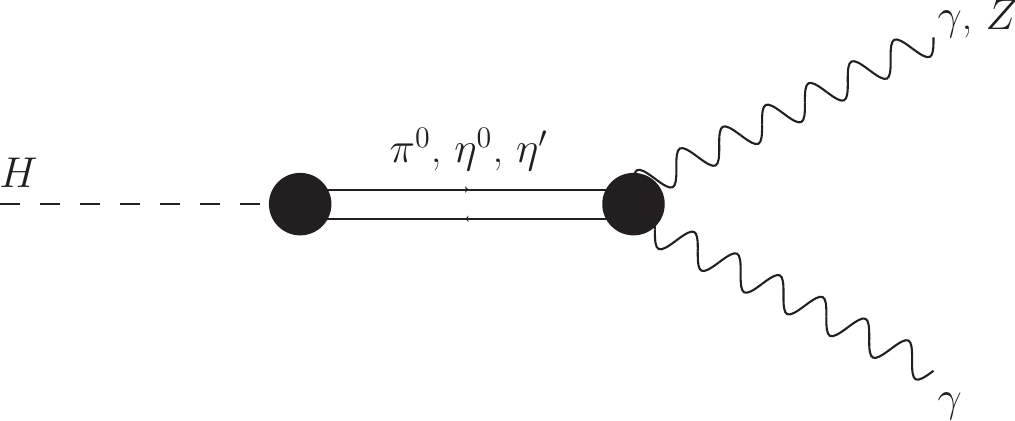}
\caption{$H\to V\gamma$ ($V=Z$, $\gamma$) decay mediated by the $\pi^0$, $\eta^0$ and $\eta^\prime$ mesons.} \label{trans}
\end{center}
\end{figure}
 

At low energies, the light quarks form bound states and the $H\to V\gamma$ ($V=\gamma$, $Z$) decay is mediated by $\pi^0$, $\eta^0$ and $\eta^\prime$ mesons as shown in Fig. \ref{trans}. These particles are described by the states:
\begin{equation}
\label{pion}
\pi^0=\frac{\overline{u}u-\overline{d}d}{\sqrt{2}},
\end{equation}
\begin{equation}
\label{eta}
\eta^0=\frac{\overline{u}u+\overline{d}d-2\overline{s}s}{\sqrt{6}},
\end{equation}
\begin{equation}
\label{etap}
\eta^\prime=\frac{\overline{u}u+\overline{d}d+\overline{s}s}{\sqrt{3}}.
\end{equation}

Moreover, the coupling of the Higgs boson with the light quarks is given as 

\begin{equation}
\label{ LagC}
\mathcal{L}=-\frac{g}{2m_W}\Big(m_u \overline{u}u+ m_d \overline{d}d+ m_u \overline{s}s \Big)H,
\end{equation}

Therefore, the amplitude for the decay $H\to \gamma\gamma$ mediated by a meson can be expressed as follows

\begin{equation}
\label{ amp}
\mathcal{M}(H\to\gamma\gamma)=-\frac{g }{2m_W  (m_H^2-m^2_{P^0})}\bra{0} m_u \overline{u}u+ m_d \overline{d}d+ m_u \overline{s}s \ket{P^0}A_{P^0}^{\mu\nu}\epsilon^\mu(p_1)\epsilon^\nu(p_2),
\end{equation}
where $P^0=\pi^0$, $\eta$, $\eta^\prime$ and $A^{\mu\nu}_{P^0}$ correspondes to the $P^0\to\gamma\gamma$ amplitude. It is well known that the under parity transformations ($P$) that
\begin{align}
\label{}
    P\ket{P^0}&=-\ket{P^0},   \\
    P\ket{0}&=\ket{0}, \\
    P\overline{\psi}\psi P^{-1}&=\overline{\psi}\psi.
\end{align}
Therefore,  
\begin{equation}
\bra{0}P^{-1}P\big(m_u \overline{u}u+ m_d \overline{d}d+ m_u \overline{s}s \big)P^{-1}P\ket{\beta}=-\bra{0} m_u \overline{u}u+ m_d \overline{d}d+ m_u \overline{s}s \ket{\beta}=0
\end{equation}
and
\begin{equation}
\mathcal{M}(H\to\gamma\gamma)=0.
\end{equation}
Therefore, the light fermions do not contribute to the $H\to \gamma\gamma$ decay. This conclusion stands in contrast to those reached in Refs. \cite{ELLIS1976292, Cahn:1978nz, Bergstrom:1985hp, Decker:1991dz, BhupalDev:2013xol, Spira:1991tj, Bonciani:2015aa, Gehrmann:2015aa, PhysRevD.42.3760, PhysRevD.47.1264, DJOUADI1993255, DJOUADI1991187, Melnikov_1993, FLEISCHER2004294, DJOUADI199817}, which suggest that the contributions from light quarks can be neglected as they are insignificant. From a similar argument, the light quarks do not contribute to the $H\to Z\gamma$ decay either.

\section{conclusion}

In this note, we demonstrated that the contributions from the bottom, charm quark, and lepton to the decays $H\to\gamma\gamma$ and $H\to Z\gamma$ are significantly smaller than those arising from the top quark and $W$-loops. Therefore, as pointed out in the literature, these contributions can be disregarded. In contrast, the contributions from the up, down, and strange quarks are zero rather than negligible.

\begin{acknowledgments}
This work  was supported by UNAM Posdoctoral Program (POSDOC). We also acknowledge support from  Sistema Nacional de Investigadores (Mexico). 
\end{acknowledgments}

\bibliography{BiblioH}
\end{document}